\documentstyle[e-e-ijmpa,twoside,epsfig]{article}

\renewcommand{\thefootnote}{\fnsymbol{footnote}}      %USE SYMBOLIC FOOTNOTE

\def\beq{\begin{equation}}
\def\eeq{\end{equation}}
\def\eeq{\end{equation}}
\def\bea{\begin{eqnarray}}
\def\eea{\end{eqnarray}}
\def\dsl{\not\!\partial}
\begin{document}

\normalsize\textlineskip

% The first line has to be % for the xxx version. The second line for the
% proc version

%\pagestyle{empty}
{\vbox{\hbox{\hspace*{98mm} SLAC-PUB-8344}\hbox{\hspace*{98mm} January 2000}}}

\title{Experimental Signatures of Split Fermions in Extra Dimensions%
\footnote{Work supported by the Department of Energy under contract 
DE-AC03-76SF00515}
}

\author{Yuval Grossman%
\footnote{Based on work with N. Arkani-Hamed and M. Schmaltz.\cite{AGS}}}

\address{Stanford Linear Accelerator Center \\
Stanford University, Stanford, CA 94309 }

\maketitle\abstracts{
The smallness and hierarchy of the fermion parameters could be explained in 
theories with extra dimensions where doublets and singlets 
are localized at slightly separated points.
Scattering cross sections for collisions of such fermions 
vanish exponentially at energies high enough to
probe the separation distance. This is because the separation puts a lower
bound on the attainable impact parameter in the collision. 
The NLC, and in particular
the combination of the $e^+e^-$ and $e^-e^-$ modes,
can probe this scenario, even if the
inverse fermion separation is of order tens of TeVs. 
}

\setcounter{footnote}{0}
\renewcommand{\thefootnote}{\alph{footnote}}

\vspace*{1pt}\textlineskip

%%%%%%%%%%%%%%%%%%%%%%
\section{Introduction}
%%%%%%%%%%%%%%%%%%%%%%

The smallness and hierarchy of the fermion
masses and mixing angles are the most puzzling features 
of fermion parameters. They suggest that there exist a more fundamental
theory that generate these properties in a natural way.
Traditionally, (spontaneously broken) flavor symmetries were assumed.
Recently, with the developments in constructing
models based on compact ``large'' dimensions, new solutions which 
exploits the new space in the extra dimensions
has been proposed.\cite{AD}

One particularly interesting framework is due to Arkani-Hamed and 
Schmaltz (AS).\cite{AS}
The idea is to separate the various fermion fields in the extra dimension.
Consider for example a model where the SM gauge and Higgs fields
live in the bulk of one extra compact dimension of radius 
of order TeV$^{-1}$
while the SM fermions are localized at different positions with
narrow wavefunctions in the extra dimension. (The gauge fields
may also be confined to a brane of thickness of about an inverse TeV
in much larger extra dimensions. Then the fermions would be stuck
to thin parallel ``layers'' within the brane.)
This separation of the fermion fields 
suppresses the Yukawa couplings. The reason is that the Yukawa couplings
are proportional to the direct couplings between the two fermion 
fields (e.g., $e_L$ and $e_R$ for the electron Yukawa coupling). 
When fermion fields are separated
the direct couplings between them is exponentially suppressed by the
overlap of their wavefunctions.
Actually, any interaction in this setup is proportional to the
overlap of the wavefunctions of the fields involved.
For example, higher dimensional
operators such as $QQQL$ which lead to proton decay
can be suppressed to safety by separating the quarks and lepton fields.

In this talk we present model independent experimental
signature of this scenario which follows simply from locality in the extra
dimensions: At energies above a TeV, the large angle scattering
cross section for fermions which are separated in the extra dimensions
falls off exponentially with energy.
This is understood from the fact that the fermion
separation in the extra dimension implies a minimum impact parameter
of order TeV$^{-1}$. At energies corresponding to shorter distances
the large angle cross section falls off exponentially because the particles
``miss'' each other. The amplitude involves a Yukawa propagator for 
the exchanged gauge boson where the four dimensional momentum transfer 
acts as the mass in the exponential. More precisely, any $t$ and $u$ channel 
scattering of split fermions has an exponential suppression.
However, $s$ channel exchange is time-like, and therefore
the fermion separation in space does not force an exponential suppression.
Nevertheless, $s$ channel processes also lead to interesting signatures
as the interference of the SM amplitude with Kaluza Klein (KK) 
exchange diagrams depends on the fermion separation.

%The remainder of this paper is structured as follows: Section 2 reviews
%the basic setup and explains how quark lepton separation suppresses
%proton decay. In Section 3 we develop the necessary formulae to calculate
%scattering cross sections in our framework. In Section 4 we apply the results
%of section 3 to different physical systems (deep inelastic scattering,
%$e^+e^-$ and $\mu^+\mu^-$ scattering) and show the reach and physics potential
%of various colliders. Section 5 contains final discussion.

%%%%%%%%%%%%%%%%%%%%%%%%%
\section{The AS Framework}
%%%%%%%%%%%%%%%%%%%%%%%%%

In this section we briefly describe the AS framework.\cite{AS}
Our starting point is the observation that
simple compactifications of higher dimensional theories
typically do not lead to chiral fermions.
The known mechanisms which do lead to chiral spectra usually
break translation invariance in the extra dimensions and
the chiral fermions are localized at special points in the compact space.
Examples include twisted sector fermions stuck at orbifold
fixed points in string theory, chiral states from intersecting D-branes,
or zero modes trapped to defects in field theory. Given that fermions
generically are localized at special points in the extra dimensions we
are motivated to consider the possibility of having different locations
for the different SM fermions. In such a scenario locality in the higher
dimensions forbids direct couplings between
fermions which live at different places. 

To be specific we concentrate on the field theoretical model 
of AS. Consider a theory with one infinite spatial dimension.
We introduce a scalar field ($\Phi$) where we assume its expectation
value to have the shape of a domain wall transverse to the extra
dimension centered at $x_5=0$. Moreover, we use the
linear approximation for the vev
in the vicinity of zero $\langle \Phi (x_5) \rangle = 2 \mu^2 x_5$.
The action for a five dimensional fermion $\Psi$ coupled to the background
scalar is then
\beq
S = \int {\rm d}^4{\bf x}\ {\rm d}x_5\,
\overline\Psi[i\!\!\dsl_4+i\gamma^5\partial_5+\Phi(x_5)] 
\Psi\ .
\label{single5}
\eeq
Performing the KK reduction one find the left handed fermion zero mode
wavefunction
\beq
\psi_L(x_5)=N 
e^{-\mu^2{x_5}^2}\,, \qquad N^2=\frac{\mu}{\sqrt{\pi/2)}} \,.
\eeq
One also finds a right handed fermion zero mode, but for 
an infinite $x_5$ this mode cannot be normalizable.

We can easily generalize Eq. (\ref{single5}) to the case of several
fermion fields. We simply couple all 5-d Dirac fields to the same
scalar $\Phi$
\beq
S = \int {\rm d}^5x\, \sum_{i,j} \bar\Psi_i[i\!\dsl_5+ 
\lambda\Phi(x_5)-m]_{ij} \Psi_j\,.
\label{multi5}
\eeq
Here we allowed for general Yukawa couplings $\lambda_{ij}$ and
also included masses $m_{ij}$ for the fermion fields. Mass terms
for the five-dimensional fields are allowed by all the symmetries
and should therefore be present in the Lagrangian.
In the case that we will eventually be
interested in -- the standard model -- the fermions carry gauge charges.
This forces the couplings $\lambda_{ij}$ and $m_{ij}$
to be block-diagonal, with mixing only between fields with identical
gauge quantum numbers. 

When we set $\lambda_{ij}=\delta_{ij}$,
$m_{ij}$ can be diagonalized with eigenvalues $m_i$,
and the resulting wave functions are Gaussians with
offset centers
\beq
\psi_L^i(x_5)=N 
e^{-\mu^2(x_5-x_5^i)^2}\,, \qquad x_5^i = {m_i \over 2 \mu^2}\,.
\eeq
When $\lambda_{ij}$ is not proportional to the unit matrix,
$m_{ij}$  and $\lambda_{ij}$ cannot be 
diagonalized simultaneously. Nevertheless, even in that case
the picture is not dramatically 
altered. While the wave functions are no longer Gaussians, 
they keep the important property to our purpose, namely,
they are still narrowly peaked at different points.
Actually, the wave functions can be approximated
by Gaussians with slowly varying 
$x_5^i$. Near the peaks they resemble Gaussians with
$x_5^i=m_i/ 2 \mu^2$ where $m_i$ are the 
eigenvalues of $m_{ij}$. 
At the tails, they look like Gaussians with
$x_5^i=m_{ii}/ 2 \mu^2$ where $m_{ii}$ are the 
diagonal values of $m_{ij}$ in the basis where $\lambda_{ij}$ is diagonal. 

Eventually, we need to be working with a finite $x_5$. 
While most of the above results hold also for finite volume, there
are some problems. First, in the simplest compactification
scenarios an anti-domain wall has to 
be introduced. This is a problem since the domain wall and the
anti-domain wall will prefer to annihilate each other. Namely, the ground
state is flat. 
Without getting into details, we only
mention that more complicated models can be constructed where 
anti-domain wall does not have to be introduced,\cite{DvSh}
or when the domain wall/anti-domain wall
configuration is stable.\cite{ASp}
The second problem is that for finite volume
the right handed zero modes
are normalizable and they are localized at the anti-domain wall or at the
boundaries of the extra dimension. Clearly, since we did not observe
them, such ``mirror'' fermions
have to be much heavier then the 
SM quarks and leptons. 
Moreover, the electroweak precision measurements,\cite{PDG} and
in particular the bound from the $S$ parameter, disfavor mirror 
families. This constrain can be avoided if there are extra, negative
contribution to the $S$ parameter. While the exact new contribution form the
KK modes in the AS model has not been calculated,
some of them do have a negative contribution.\cite{AS}

Now we are in the position to calculate the Yukawa coupling. 
We demonstrate it for the electron case.
In the
five dimensional theory the Yukawa interaction is giving by
\beq \label{double5}
S_Y=  \int d^5x \kappa H\bar L E\,.
\eeq
We assume that the relevant 
massless zero mode from $L$ is localized at
$x_5=0$ and that from $E$ at $x_5=r$. 
For simplicity, we will assume
that the Higgs is delocalized inside the wall.

We now determine what effective four-dimensional interactions between
the light fields results from the Yukawa coupling in eq. (\ref{double5}).
To this end we replace $L$, $E$ and the Higgs field $H$ by their lowest
Kaluza-Klein modes. 
We obtain for the Yukawa coupling
\beq
S_Y = \int {\rm d}^4{\bf x}\,\, \kappa\, h({\bf x}) l({\bf x}) e
({\bf x})\ \int {\rm d}x_5\ \psi_l(x_5)\ \psi_{e}(x_5)\ .
\label{yuk4}
\eeq
The zero-mode wave functions
for the lepton doublet and singlet, $\psi_l(x_5)$ and $\psi_{e}(x_5)$,  
are Gaussian centered at $x_5=0$  and
$x_5=r$ respectively. 
%For simplicity we assume that
%the Higgs field has an $x_5$-independent wave function.
Overlap of Gaussians is itself a Gaussian and we find
\beq
\int {\rm d}x_5\ \psi_l(x_5)\ \psi_{e}(x_5) = \frac{\sqrt{2}\mu}{\sqrt{\pi}}
\int {\rm d}x_5\ e^{-\mu^2x_5^2} e^{-\mu^2(x_5-r)^2}
= e^{-\mu^2r^2/2}\ .
\eeq
We found that the Yukawa coupling is highly suppressed once $r$ is not much 
smaller than $\mu^{-1}$.
We finally note that a concrete model that reproduced the observed
quark and lepton parameters has been worked out in.\cite{MiSch}

%%%%%%%%%%%%%%%%%%%%%%%%%%%%%%%%%%
\section{Scattering of fermions localized at different places}
%%%%%%%%%%%%%%%%%%%%%%%%%%%%%%%%%%

We now discuss scattering of fermions localized at different places
at the extra dimension.\cite{AGS}
%Here we follow closely the discussion in 
%\cite{AGS} where more details can be found.
%
Let us imagine colliding fermions which are localized at two
different places in a circular extra dimension of radius $R$.
For concreteness we consider the 
scattering of right handed electrons on left handed electrons.
In the context of our model there are three potentially relevant mass
scales for this collision: the momentum transfer of the $t$-channel
scattering $\sqrt{-t}$,
the inverse of the quark-lepton separation $d^{-1}$ which we take
to be of order of the inverse thickness $R^{-1}$ of 
the extra dimension, and the inverse
width of the fermion wave functions $\sigma^{-1}$. However, we
will approximate the fermion wave functions by delta functions
for the calculation. The
corrections which arise from the finite width of the wave functions
were calculated and found to be negligible for practical purposes.\cite{AGS}

To calculate the scattering through intermediate bulk gauge fields
we need the five-dimensional propagator.
In momentum space it is $(t-p_5^2-m^2)^{-1}$
where we separated out the five dimensional momentum transfer $p_5$.
As we are interested in propagation between definite positions in
the fifth dimension it is convenient to Fourier transform in the
fifth coordinate
\beq
P_d(t)= \sum_{n=-\infty}^{\infty} {e^{ind/R}
    \over t-({n/R})^2-m^2}\ ,
\label{fiveprop}
\eeq  
where $d=x_L-x_R$ and $x_L(x_R)$ is the location of $e_L(e_R)$ in the 
extra dimension. The Fourier transform is a sum and not an 
integral since momenta in the fifth coordinate
are quantized in units of $1/R$.
This propagator can also be understood in the four dimensional (4d) language
as arising from exchange of the 4d gauge boson and its infinite tower of KK
excitations.\cite{AGS}
This propagator can be simplified by performing the sum. 
We find\cite{AGS}
\beq
P_d(t)= - {\pi R\over \sqrt{-t+m^2}}\ 
  {\cosh[(d-\pi R)\sqrt{-t+m^2}]\over \sinh[\pi R\sqrt{-t+m^2}]}\,.
\label{towerprop}
\eeq
The Feynman rules for diagrams involving exchange of bulk gauge fields
are now identical to the usual four dimensional SM Feynman rules
except for the replacement of 4d gauge boson propagators by the corresponding
5d propagators. Before we proceed with calculating cross sections
we note a few properties of the above propagator.

It is easy to understand the two limits
$\sqrt{-t}\gg R^{-1}$ and $\sqrt{-t}\ll R^{-1}$. In the former
case we obtain 
\beq
P_d(t) \simeq - {\pi R\over \sqrt{-t}}\ e^{-\sqrt{-t}\,d}\,,
\eeq
which vanishes exponentially with the momentum transfer in the process
as we anticipated from five dimensional locality.
In the limit of small momentum transfer we obtain
\beq \label{small-t}
P_d(t) \simeq {1\over t-m^2} - 
       R^2\left(\frac{d^2}{2R^2}-{d \pi \over R}+\frac{\pi^2}{3}\right)\,,
\eeq
which is the four dimensional $t$-channel propagator
plus a correction term whose sign and magnitude depends on the
fermion separation. For small separation $d < \pi R\,(1-1/\sqrt{3})$ the 
correction enhances the magnitude of the amplitude, while for larger separation
it reduces it.  

It is also instructive to expand the propagator in exponentials (ignoring
the mass $m$)
\beq \label{expand}
P_d(t)=- {\pi R\over \sqrt{-t}} 
       \left( e^{-\sqrt{-t}d}+e^{\sqrt{-t}(d-2\pi R)}\right)
       \left( 1+e^{-\sqrt{-t}2\pi R}+e^{-\sqrt{-t}4\pi R}+
              \dots\right)\,,
\eeq
which can be understood as a sum of contributions from five
dimensional propagators. The two terms in the first parenthesis
correspond to propagation from $x_L$ to $x_R$ in clockwise and 
counter--clockwise directions, and the series in the other parenthesis adds
the possibility of also propagating an arbitrary number of times around the
circle. 

The expression for the $u$-channel KK-tower
propagator $P_d(u)$ is identical to eq.~(\ref{towerprop}) with the obvious
replacement $t\rightarrow u$, and $P_d(s)$ is obtained by analytic
continuation
\beq
P_d(s)= {\pi R\over \sqrt{s-m^2}}
  {\cos[(d-\pi R)\sqrt{s-m^2}]\over \sin[R\pi\sqrt{s-m^2}]}\ .
\label{schannel}
\eeq
The poles at $\sqrt{s-m^2}=n/R$ are not physical and 
can be avoided by including a finite width.

Armed with this propagator it is easy to evaluate any KK boson
exchange diagram in terms of its SM counterpart. For example, a pure
$t$ channel exchange diagram becomes
\beq
{\cal M} =  (t-m^2) P_d(t) \times {\cal M}\big|_{SM} \ ,
\eeq
where ${\cal M}\big|_{SM}$ is the SM amplitude and the factor 
$(t-m^2) P_d(t)$
replaces the SM gauge boson propagator $1/(t-m^2)$ by the 5d
propagator $P_d(t)$. 

%%%%%%%%%%%%%%%%%%%%%%%%%%%%%%%%%%%%%%%%%%
\section{Collider signatures}
%%%%%%%%%%%%%%%%%%%%%%%%%%%%%%%%%%%%%%%%%%

Having calculated the 5d propagator, the calculation of 
differential cross sections is a simple generalization of SM
results.\cite{book} 
We start by considering the predictions of our model for
high energy $e^+e^-$ or $\mu^+\mu^-$ machines.
We assume that the separation of left and right
handed fields is responsible for at least part of the suppression of the
muon and electron Yukawa couplings.
Therefore, we consider a case where
the doublet and singlet components of the charged leptons are split
by a distance $d$ in the extra dimensions.
We start by looking into a pure $t$ channel exchange 
which is (in principle) possible
at a lepton collider with polarizable beams
$l_L^+ l_R^- \longrightarrow l_L^+ l_R^-$.
To compute the differential cross section we
sum over contributions from neutral current exchange (photon and $Z$ plus
KK towers). In the formulae 
in this section we neglect $m_Z$. It is easy to 
reintroduce it, and in our numerical plots we keep it.
Happily,
each term in the sum is simply equal to the SM term times $t P_d(t)$
which can be factored so that our final expression for the 
differential cross section becomes
\beq \label{ep} 
r_\sigma^t \equiv {{d\sigma / dt} \over
{d\sigma / dt} \left|_{\rm SM} \right. }
=  \left|t\, P_d(t)\right|^2 \,,
\eeq
where $P_d(t)$ is given in eq.~(\ref{towerprop}). 
The effect of the KK tower would be seen as a dramatic
reduction of the cross section at large $|t|$.
To illustrate this point
in Fig. 1 we plot the ratio $r_\sigma^t$ of eq.~(\ref{ep}) as a function
of $t$ for $R=1$ TeV$^{-1}$ and representative values of $d$.  

% FIGURE 1
\begin{figure}[ht]
\epsfig{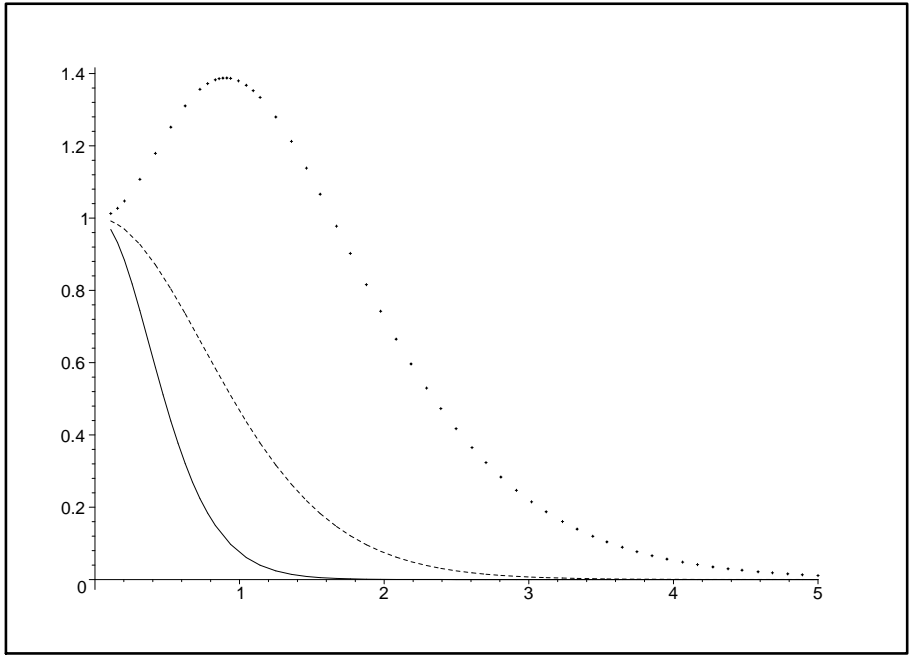}
\vspace*{-28mm}
\caption{$r_\sigma^t$ (the cross section for $t$ channel exchange
in the 5d theory normalized by the corresponding SM cross section)
as a function of $\sqrt{-t}$ in units of TeV. 
We assume $R^{-1}=1\,$TeV. The dotted, dashed and solid curves are for
separations of $d/R=1$, $\pi/2$ and $\pi$ respectively.}
\end{figure}

Experimental Signatures of Split Fermions in Extra Dimensions
We can get more information on the values on $d$ and $R$ by 
combining the above with the processes
$e^+_N e^-_N \to \mu^+_N \mu^-_N$ ($N=L$ or $R$). 
(The same considerations also
apply to scattering into quark pairs, but this case is 
more difficult to study experimentally.)
This process is a pure
$s$ channel between unseparated fermions so that
\beq \label{ee-pol}
r_\sigma^{sN} \equiv {{d\sigma / dt} \over
{d\sigma / dt} \left|_{\rm SM} \right. }
=  \left|s\, P_0(s)\right|^2 \,.
\eeq
For $\sqrt{s}$ small
compared to the inverse size of the extra dimension
the cross section is reduced independently of $d$.
An extra dimensional theory without fermion separation predicts 
$r_\sigma^{sN} < 1$ and $r_\sigma^{t} > 1$.
Thus, a measurement of $r_\sigma^{sN} < 1$ together with $r_\sigma^t < 1$
would be evidence for fermion separation in the extra dimension.

Another interesting probe of $d$ using $s$ channel has been 
suggested by Rizzo.\cite{T}
Suppose that the first KK mode has been produced and its 
mass $1/R$ measured.
The case of $d=0$ can be distinguished from $d \neq 0$ by 
looking at the cross-section at lower energies. 
In particular, for $d=0$, the first KK exchange exactly cancels 
the SM amplitude at 
$\sqrt{s} = 1/(\sqrt{2}R)$, whereas for $d \neq 0$ the cross-section can 
still be large. 
Therefore, a beam scan at energies beneath the first resonance can 
be an efficient probe of $d$.
 
Even if beam polarization is not available, one can still 
probe the nature of the extra dimensions by looking at several processes
and using angular information. Consider an unpolarized
$e^+ e^- \to \ell^+ \ell^-$ scattering. (The same holds for incoming muons.)
We get the tree level cross section
\beq \label{ee-un}
{d\sigma \over dt} = {\pi \alpha^2 \over s^2} \left[
   \left(1+{1\over16\sin^4\theta_w}\right) 
{u^2 (P_0(s)+P_0(t))^2 \over \cos^4\theta_w}
   + {t^2P_d^2(s)+s^2P_d^2(t) \over 2 \cos^4\theta_w} \right]\,.
\eeq
When $\ell=e$ both $s$ and $t$ channels are possible, while for $\ell \ne e$
only the $s$ channel is present, and in the above formula one should set 
$P_d(t) = P_0(t) = 0$. We also define, as before, the ratio of the 5d 
cross section to the
SM one as $r_\sigma^{s}$ ($r_\sigma^{st}$) for the 
$e^+ e^- \to \mu^+ \mu^-$ ($e^+ e^- \to e^+ e^-$) reaction.
In Fig.~2 we presented
$r_\sigma^{st}$ as a function of the scattering angle.
As we can see, the cross sections depend in 
a non trivial way on the separation. This is
because the helicity changing amplitude depends on $d$, while the helicity
conserving one does not. 
By looking at angular distributions, one can separate the different 
contributions, and extract both $R$ and $d$. 

% FIGURE 2
\begin{figure}[ht]
\epsfig{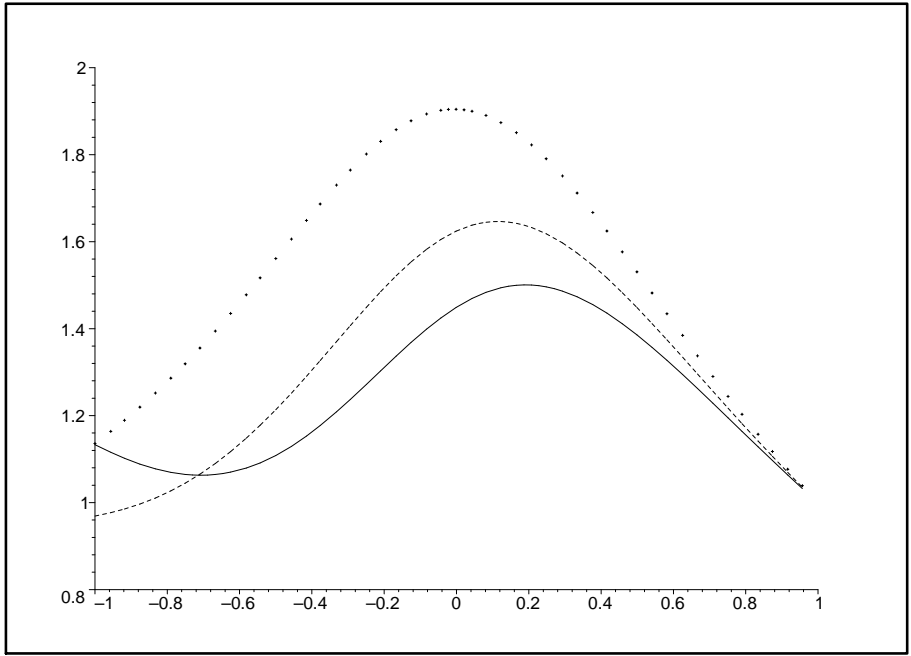}
\vspace*{-28mm}
\caption{$r_\sigma^s$ (the cross section for
$e^+ e^- \to e^+ e^-$,
in the 5d theory normalized by the corresponding SM cross section)
as a function of the scattering angle, $\cos\theta$. 
We assume $R^{-1}=4\,$TeV and $\sqrt{s}=1.5\,$TeV. 
The dotted, dashed and solid curves are for
separation of $d/R=0$, $1$ and $\pi$ respectively.}
\end{figure}

Another interesting collider mode which allows a very clean measurement
of fermion separations is $e^-e^-$ scattering. The advantage of this 
mode is that both beams can be polarized to a high degree which
allows for a clean separation of the interesting $t$ and $u$ channels from
$s$ channel.
We find for $e_L^- e_R^-$ scattering to $e^- e^-$ (summed over final
polarizations)
\beq \label{emin-emin}
r_\sigma^{+-} \equiv {{d\sigma / dt} \over
{d\sigma / dt} \left|_{\rm SM} \right.} =
{u^2 |P_d(t)|^2 + t^2 |P_d(u)|^2 \over
             u^2/t^2 + t^2/u^2}\,.
\eeq
Higher sensitivity to $d$ can be achieved by
changing the electron polarization.
Consider $e^- e^-$ scattering where one electron if left handed and the
other has polarization $p$ which can vary between $-1$ and $1$.
We find
\beq
r_\sigma^{p-} - 1 \approx 2 R^2(|t|+|u|) 
\left[{1+p\over 2}\left(
{d^2\over R^2}-{2\pi d \over R}\right) + {2\pi^2 \over 3}\right]
\eeq
Varying $p$ one could, in principle, determined both $d$ and $R$.

While an exponential suppression of the cross section would be an 
unambiguous signal of fermion separation in the extra dimension, we
can still probe $d$ if a small deviation of $r_\sigma$
from unity is found. The sensitivity can estimated from
eq.~(\ref{small-t}). Assuming maximum separation, $d=\pi R$, 
there is a reduction in the cross-section 
($r_\sigma^t < 1$), and we obtain
a sensitivity 
\beq \label{t-sens}
R \le \sqrt{3 \, \Delta r_\sigma^a\over \pi^2 Q^2}\,,
\eeq
where $\Delta r_\sigma^a$ is the combined theoretical and experimental error
on $r_\sigma^a$. For $d=0$ one should find $r_\sigma^t>1$ with 
a factor of $\sqrt{2}$ higher sensitivity.
Assuming $\Delta r_\sigma \approx 1\%$ and using eq.~(\ref{t-sens})
we conclude that we get sensitivity  
down to $R \approx (27\,$TeV$)^{-1}$ at a $1.5\,$TeV linear collider.

%%%%%%%%%%%%%%%%%%%%%%%%%%%%%%%%%
\section{Conclusion}
%%%%%%%%%%%%%%%%%%%%%%%%%%%%%%%%%

Fermion separation in extra dimension 
is a useful model building tool. It can
explain the smallness and hierarchy of the 
Yukawa couplings in the SM and suppress proton decay in models
with low fundamental scale.
A model independent prediction of this framework is that the 
space-like exchange amplitudes between 
split fermions falls off exponentially.
If the inverse 
size of the extra dimension is not much larger then 10 TeV, we can see
this fall off.
The NLC, and in particular the combination of the $e^+e^-$ and $e^-e^-$
options, is very promising in this respect.

%%%%%%%%%%%%%%%%%%%%%%%%
%\acknowledgements
%%%%%%%%%%%%%%%%%%%%%%%%
%We thank Stan Brodsky, Hooman Davoudiasl, Lance Dixon, Hitoshi Murayama
%and Tom Rizzo for useful discussions. N.A.-H. is supported by DOE under
%contract DE-AC03-76SF00098 and by NSF under contract PHY-95-14797. Y.G.
%and M.S. are supported by the Department of Energy under contract
%DE-AC03-76SF00515.

\def\pl#1#2#3{{\it Phys. Lett. }{\bf B#1~}(19#2)~#3}
\def\zp#1#2#3{{\it Z. Phys. }{\bf C#1~}(19#2)~#3}
\def\prl#1#2#3{{\it Phys. Rev. Lett. }{\bf #1~}(19#2)~#3}
\def\rmp#1#2#3{{\it Rev. Mod. Phys. }{\bf #1~}(19#2)~#3}
\def\prep#1#2#3{{\it Phys. Rep. }{\bf #1~}(19#2)~#3}
\def\pr#1#2#3{{\it Phys. Rev. }{\bf D#1~}(19#2)~#3}
\def\np#1#2#3{{\it Nucl. Phys. }{\bf B#1~}(19#2)~#3}
\def\epjc#1#2#3{{\it Eur. Phys. J.}{\bf C#1~}(19#2)~#3}

\nonumsection{References}

\end{document}